\PassOptionsToPackage{table}{xcolor}
\documentclass[twocolumn,prd,preprintnumbers,numbers,sort&compress,nofootinbib,showpacs,colorlinks,citecolor=blue,amsmath,amssymb,aps,superscriptaddress]{revtex4-2}

\usepackage{graphicx}% Include figure files
\usepackage{dcolumn}% Align table columns on decimal point
\usepackage{bm}% bold math
\usepackage{xcolor}
\usepackage{lipsum} 
\usepackage{float} 
\usepackage{CJKutf8}

\newcommand{\be}{\begin{equation}}
\newcommand{\ee}{\end{equation}}
\newcommand{\bw}{\begin{widetext}}
\newcommand{\ew}{\end{widetext}}
\newcommand{\ba}{\begin{aligned}}
\newcommand{\ea}{\end{aligned}}
\newcommand{\bes}{\begin{equation*}}
\newcommand{\ees}{\end{equation*}}
\newcommand{\bea}{\begin{eqnarray}}
\newcommand{\eea}{\end{eqnarray}}
\newcommand{\dd}{\text{d}}

\newcommand{\hmC}{\hat{\mathcal{C}}}

\newcommand{\fR}{\mathfrak{R}}
\newcommand{\hfR}{\hat{\mathfrak{R}}}
\newcommand{\tFD}{\mathcal{T}_\mathrm{FD}}
\newcommand{\pcl}{pseudo-$C_{\ell}\,$}

\newcommand{\beq}{\begin{equation}}
\newcommand{\eeq}{\end{equation}}
\newcommand{\exclude}[1]{}

\definecolor{Orange}{rgb}{1.0,0.5,0.15}
\definecolor{Blue}{rgb}{0,0.08,0.65}
\definecolor{Red}{rgb}{0.65,0.08,0.05}
\definecolor{Green}{rgb}{0.15,0.45,0.25}
\definecolor{Pink}{rgb}{1.0,0.05,0.5}
\definecolor{bubbles}{rgb}{0.91, 1.0, 1.0}
\definecolor{aquamarine}{rgb}{0.5, 1.0, 0.83}
\definecolor{bubblegum}{rgb}{0.99, 0.76, 0.8}
\definecolor{bluebell}{rgb}{0.74, 0.74, 0.92}
\definecolor{dollarbill}{rgb}{0.72, 0.93, 0.6}
\definecolor{cred}{RGB}{238,28,37}

\newcommand{\ubc}{Department of Physics and Astronomy, University of British Columbia, 6224 Agricultural Road, Vancouver, British Columbia V6T 1Z1, Canada}

\newcommand{\bochum}{German Centre for Cosmological Lensing, Astronomisches Institut, Ruhr-Universit\"at Bochum, Universit\"atsstrasse 150, 44780, Bochum, Germany}

\newcommand{\nagoyaU}{Graduate School of Science, Nagoya University, Furocho, Chikusa-ku, Nagoya, Aichi, 464-8602, Japan}

\newcommand{\cea}{Universit\'e Paris-Saclay, IPHT, DRF-INP, UMR 3681, CEA, Orme des Merisiers Bat 774, 91191 Gif-sur-Yvette, France}

\newcommand{\cgpqi}{Center for Gravitational Physics and Quantum Information, Yukawa Institute for Theoretical Physics, Kyoto University, Kyoto 606-8502, Japan}

\newcommand{\inns}{Universit\"at Innsbruck, Institut f\"ur Astro- und Teilchenphysik, Technikerstrasse 25/8, 6020 Innsbruck, Austria}

\newcommand{\ifae}{Institut de F\'isica d’Altes Energies (IFAE), The Barcelona Institute of Science and Technology, Campus UAB, 08193 Bellaterra (Barcelona), Spain}

\newcommand{\bonn}{Argelander-Institut f\"ur Astronomie, Universit\"at Bonn, Auf dem H\"ugel 71, D-53121 Bonn, Germany}

\newcommand{\kipac}{Kavli Institute for Particle Astrophysics and Cosmology, Stanford University, Stanford, California 94305, USA}
\newcommand{\slac}{SLAC National Accelerator Laboratory, Menlo Park, California 94025, USA}
\newcommand{\kmi}{Kobayashi-Maskawa Institute for the Origin of Particles and the Universe (KMI), Nagoya University, Nagoya, 464-8602, Japan}
\newcommand{\unibo}{Dipartimento di Fisica e Astronomia ``Augusto Righi''-Alma Mater Studiorum Università di Bologna, via Piero Gobetti 93/2, I-40129 Bologna, Italy}
\newcommand{\inaf}{Istituto Nazionale di Astrofisica (INAF)-Osservatorio di Astrofisica e Scienza dello Spazio (OAS), via Piero Gobetti 93/3, I-40129 Bologna, Italy}
\newcommand{\infn}{Istituto Nazionale di Fisica Nucleare (INFN)-Sezione di Bologna, viale Berti Pichat 6/2, I-40127 Bologna, Italy}

\newcommand{\warsaw}{Center for Theoretical Physics, Polish Academy of Sciences, Aleja Lotnik\'ow 32/46, 02-668 Warsaw, Poland}

\newcounter{FFcounter}

\newcounter{LVWcounter}

\newcommand{\yanza}[1]{\protect\textcolor{black}{#1}}
\newcommand{\gsm}[1]{\protect\textcolor{black}{#1}}

\newcounter{SGcounter}

\newcounter{RDcounter}

\usepackage{hyperref}

\begin{document}

%\preprint{APS/123-QED}

\title{Consistency test of the large-scale structure with Bernardeau-Nishimichi-Taruya transform of the KiDS-Legacy survey}

\author{Shiming Gu (\begin{CJK}{UTF8}{gkai}{顾时铭}\end{CJK})}
 \email{gsm@phas.ubc.ca}
      \affiliation{\ubc}
      \affiliation{\bochum}

\author{Ziang Yan (\begin{CJK}{UTF8}{gkai}{颜子昂}\end{CJK})}
\email{yan.ziang.h1@f.mail.nagoya-u.ac.jp}
      \affiliation{\bochum}
      \affiliation{\nagoyaU}
      \affiliation{\kmi}

\author{Ludovic van Waerbeke}%
 \email{waerbeke@phas.ubc.ca}
      \affiliation{\ubc}

\author{Francis Bernardeau}
      \affiliation{\cea}
      \affiliation{\cgpqi}

\author{Hendrik Hildebrandt}
      \affiliation{\bochum}

\author{Angus H. Wright}
      \affiliation{\bochum}

\author{Maciej Bilicki}
      \affiliation{\warsaw}

\author{Christos Georgiou}
      \affiliation{\ifae}

\author{Shun-Sheng Li (\begin{CJK}{UTF8}{gkai}{李顺生}\end{CJK})}
      \affiliation{\kipac}
      \affiliation{\slac}

\author{Laila Linke}
      \affiliation{\inns}

\author{Lauro Moscardini}
      \affiliation{\unibo}
      \affiliation{\inaf}
      \affiliation{\infn}

\author{Robert Reischke}
      \affiliation{\bonn}
      \affiliation{\bochum}

\author{Benjamin St\"olzner}
      \affiliation{\bochum}

\date{\today}

\begin{abstract}
We perform the first $k$-cut cosmic shear analysis of the KiDS-Legacy survey. This method uses the Bernardeau-Nishimichi-Taruya (BNT) transform to construct weak-lensing kernels that are more localised than conventional ones, and remove information from selected physical scales while retaining the constraining power of the targeted range. Removing the scale of $k \geq 0.33~\mathrm{Mpc}^{-1}$ from the KiDS-Legacy pseudo-$C_\ell$ data vector, and using a covariance matrix whose Gaussian component is computed from the theoretical data vector, we find $S_8 = 0.798_{-0.045}^{+0.045}$. This agrees with both the fiducial KiDS-Legacy bandpower result and our no-$k$-cut pseudo-$C_\ell$ posterior to within $0.1\sigma$, indicating no significant bias from nonlinear astrophysical feedback at the precision of KiDS-Legacy.
We also study the case in which the Gaussian covariance is computed from the observed data vector. In this setup, the same scale cut of $k < 0.33~\mathrm{Mpc}^{-1}$ gives a much lower $S_8=0.717_{-0.046}^{+0.047}$. Further $k$-cut tests reveal a mild scale-dependent trend, with larger physical scales preferring lower $S_8$ values and a maximum low- versus high-$k$ deviation of $1.80\sigma$. Mock tests show that this behaviour is not produced by the covariance prescription or data vector alone, but may arise from their interplay. These results show that BNT $k$-cuts provide both a mitigation strategy for nonlinear systematics and a diagnostic of weak-lensing inference pipelines.

\end{abstract}

\maketitle

\section{Introduction}

\gsm{Weak gravitational lensing provides one of the most direct late-time probes of the amplitude of matter clustering, usually parametrised} as $S_8=\sigma_8(\Omega_m/0.3)^{0.5}$, where $\sigma_8$ is the amplitude of density fluctuations in the universe on a scale of $8 ~\mathrm{Mpc}/h$ and \gsm{$\Omega_m$ is the present-day matter density parameter. Weak-lensing} measurements \gsm{infer this clustering amplitude statistically} \cite{vanWaerbeke2000WL,Kaiser2000WL,Bacon2000WL,Wittman2000WL} \gsm{from correlated changes in galaxy shapes}  \cite{Blandford1991WL,Kaiser1992Shear}. This technique, \gsm{known as \textit{cosmic shear}, has become} a powerful cosmological probe of the distribution and growth of matter, enabling stringent tests of dark matter and dark energy models \cite{Bartelmann2001WL}.
 
In recent years, a growing number of large imaging surveys have employed weak lensing to map the matter density field with unprecedented precision. These measurements have revealed tensions in the inferred value of $S_8$ when compared to predictions from the cosmic microwave background under the Lambda cold dark matter ($\Lambda$CDM) model, \gsm{the standard cosmological model, which motivates the research of possible modelling systematics and extensions to the standard cosmological model. For example,} KiDS-1000 analysis reported a relatively low clustering amplitude, $S_8 = 0.759^{+0.024}_{-0.021}$, corresponding to a $\sim3\sigma$ offset from Planck CMB predictions in simple $S_8$ comparisons \cite{Asgari2021KiDS}.  Nevertheless, the most recent KiDS-Legacy cosmic shear results \cite{Wright2025KiDSLegacy,Stolzner25KiDSLegacy}, based on the full survey footprint and improved photometric redshift estimation, yield $S_8 = 0.813^{+0.018}_{-0.018}$ (COSEBIs marginal mean + CI)\footnote{Please note that the official KiDS-Legacy cosmic shear paper quoted the fiducial COSEBIs result as their main result. However, as we are using pseudo-$C_\ell$, whose data vector structure is closer to the bandpowers, we will mostly quote their bandpower results $S_8 = 0.797^{+0.023}_{-0.024}$ in this paper.}, consistent with Planck at the $0.73\sigma$ level, illustrating how expanded datasets and, more importantly, a different methodology can materially change the inferred level of tension.

In addition to uncertainties in photometric redshifts, one of the most widely discussed explanations for the $S_8$ tension involves effects in the modelling of the nonlinear regime \cite{Amon22S8ANL}.
Within the standard \gsm{picture of cosmic structure formation, linear scales are comparatively insensitive to nonlinear late-time processes and therefore provide a clean reference for testing the consistency of late-time clustering measurements} \citep{BBKS}. \gsm{Many proposed explanations of the $S_8$ tension instead act preferentially on nonlinear scales or through late-time structure growth, including unexpectedly} strong baryonic feedback \cite{Gu2023HMF,Preston23ANL}, late-time scalar fields \cite{White22DESILRGsigma8,Lin24latetimeS8}, and exotic dark matter models \cite{Naido24DarkMatterS8,Gu2025BNT1}. However, these explanations typically require either non-minimal extensions of the standard cosmological model or baryonic feedback strengths that are disfavoured by independent astronomical constraints \cite{VanDaalen20GalaxyonPk}. As a result, despite their ability to improve the fit to certain observational datasets, these \gsm{possibilities motivate a direct test of whether the inferred clustering amplitude depends on physical scale.}

However, even if such models could explain why weak lensing measurements yield a lower $S_8$ value, they generally affect the cosmic large-scale structure differently than changes in standard cosmological parameters such as the matter density $\Omega_m$ and the scalar fluctuation amplitude $A_s$. For example, some models recover the Planck $S_8$ by suppressing power in the nonlinear regime \cite{Amon22S8ANL,Gu2023HMF}. Weak lensing, however, traces all matter along the line of sight, so correlations measured at given angular scales receive contributions from a wide range of physical scales. This well-known projection effect limits the effectiveness of nonlinear corrections. To address this limitation, the Bernardeau-Nishimichi-Taruya (BNT) transform \cite{Bernardeau2014BNT} was proposed. The BNT effectively localises cosmic shear tomography to a narrower redshift range, and when combined with a well-defined angular-scale filter ($\ell$-cut in harmonic space), it can nearly serve as a 3-D physical-scale cut in Fourier space. This enables more robust mitigation of uncertainties associated with nonlinear modelling of the three-dimensional matter power spectrum. In our previous works \cite{Gu2025BNT1,Gu26BNT2}, we developed a BNT-based cosmic shear analysis pipeline, demonstrated that the BNT transform can serve as a targeted solution for nonlinear systematics, and proposed a BNT-based consistency test for modelling uncertainties within the posterior sampling process. 

From the comparison of \citet{Taylor21BNTDESSV} and \citet{Gu2025BNT1}, we also find that it is much better to use \gsm{the} BNT transform in the harmonic $\ell$-space rather than the configuration space to target the $k$-cut information more effectively. However, it is generally challenging to calculate $C_{\ell}$ on the weighted sky (for example, partially covered sky with anisotropic lensing weights). The KiDS cosmic shear analyses \citep[][]{Asgari2021KiDS, Wright2025KiDSLegacy} use the bandpower, that is, measuring real-space correlation functions first, then transferring to $\ell$-space with predefined bandpower window functions. In this work, we use the pseudo-$C_{\ell}$ method \citep[PCL, ][]{Alonso_2018}, that is, directly taking \gsm{a} harmonic \gsm{transform} from the weighted shear maps, then correcting for the mode-coupling induced by the weights. Compared to catalogue-based correlation functions, this method is relatively faster, especially when estimating the covariance \citep{Garc_a_Garc_a_2019}. In addition, one can make cleaner scale cuts with pseudo-$C_{\ell}$.

\gsm{In this work, we perform the first BNT k-cut analysis of the KiDS-Legacy cosmic shear data. We use physically selected scale ranges to test the consistency of the inferred $S_8$.} The structure of the paper is organised as follows. In Section~\ref{sec:BNT3_theory}, we introduce the theoretical foundations of weak gravitational lensing, pseudo-$C_\ell$ estimators, and the BNT transform. In Section~\ref{sec:BNT3_data}, we describe the data, the construction and manipulation of the data vector, the modelling procedure, and the pipeline for inferring cosmological parameters from the observables. In Section~\ref{sec:BNT3_results}, we present BNT-based $S_8$ constraints derived from different physical scale selections and discuss their cosmological implications. Finally, Section~\ref{sec:BNT3_conclusion} summarises our findings and provides concluding remarks.

\section{Theoretical modelling}
\label{sec:BNT3_theory}

\subsection{Power spectra}
\label{subsec:IIA}

The information properties of the cosmic large-scale structure are often described by its two-point statistics. At the second order, the matter density field is usually characterised by the power spectrum $P(k;z)$:
\be
\langle \delta(\mathbf{k};z) \, \delta^*(\mathbf{k}';z) \rangle = (2\pi)^3 \delta_D(\mathbf{k} - \mathbf{k}') P(k;z).
\ee
where $\delta_D$ is the Dirac delta function. In weak lensing observation, we probe projected fields rather than a full 3D matter distribution. The cosmic shear signal traces the integrated matter fluctuations along the line of sight, with its corresponding observable, the cosmic shear angular power spectrum $C_\ell$, which can be decomposed to the contribution of the cosmic shear, galaxy intrinsic shape (or intrinsic alignment, IA), and their cross-correlations:
\be
C^{i,j} \equiv C^{i,j}_{\rm GG} + C^{i,j}_{\rm GI} + C^{i,j}_{\rm IG} + C^{i,j}_{\rm II}.
\label{eq:C_ij_all_def}
\ee
 
The shear-shear component, $C^{i,j}_{\rm GG}$ is related to $P(k;z)$ through a line-of-sight projection weighted by the lensing efficiency and geometry \cite{Limber1954Projection}:
\be
C_{\rm GG}^{i,j}(\ell) = \int\dfrac{d\chi}{f_K(\chi)^2} W^i_\gamma(\chi)W^j_\gamma(\chi)P\left(k=\dfrac{\ell+1/2}{\chi};z(\chi)\right);
\label{Cell_original}
\ee
where $\chi$ is the radial comoving distance, $f_K(\chi)$ is the angular diameter distance at $\chi$\footnote{Throughout the paper, we assume a spatially flat Universe with zero spatial curvature, $K=0$.}:
\be
f_K(\chi)=
\begin{cases}
\dfrac{1}{\sqrt{K}}\;\sin\!\big(\sqrt{K}\,\chi\big), & K>0\ \text{(closed)},\\[6pt]
\chi, & K=0\ \text{(flat)},\\[6pt]
\dfrac{1}{\sqrt{-K}}\;\sinh\!\big(\sqrt{-K}\,\chi\big), & K<0\ \text{(open)},
\end{cases}
\ee
and $W_\gamma^i(\chi)$ is the lensing kernel computed by the source redshift distribution of the $i^{\mathrm{th}}$ tomographic bin of a given survey:
\be \label{eqn:bnt3_Wdef}
W_\gamma^i(\chi) = \dfrac{3\Omega_mH_0^2}{2c^2}\int\dd \chi' \dfrac{n_i(\chi')}{a(\chi)}\dfrac{f_K(\chi'-\chi)f_K(\chi)}{f_K(\chi')},
\ee
where $n_i(\chi(z))$ is the redshift distribution of galaxies.

In this paper, we use the nonlinear alignment model \citep[NLA, ][]{Bridle2007NLA} to model the intrinsic correlation of the galaxy shapes. It can be described by applying a redshift-dependent linear bias $C_1(z)$ to the nonlinear matter power spectrum:
\bea
P_\mathrm{GI}(k, z) &=& C_1(z) \, P_{\mathrm{nl}}(k; z) \nonumber\\
P_\mathrm{II}(k, z) &=& C_1^2(z) \, P_{\mathrm{nl}}(k, z)
\label{eq:pofkz}
\eea
where the tidal alignment bias $C_1(z)$ is always parameterised as a redshift-dependent function involving a global IA scaling amplitude $A_\mathrm{TA}$ and a power-law evolution index $\eta_\mathrm{TA}$ where we use $\eta_\mathrm{TA} = 0$ in this paper in order to be consistent with the parameter choice of \cite{Wright2025KiDSLegacy}:
\be \label{eqn:c1}
C_1(z) \equiv -A_\mathrm{TA} \, \tilde{C}_1 \, \frac{\rho_\mathrm{crit} \, \Omega_m}{D(z)} \left( \frac{1+z}{1+z_0} \right)^{\eta_\mathrm{TA}}
\ee
where $\rho_\mathrm{crit}$ is the critical density of the universe, $D(z)$ is the structure growth function, $z_0$ is a pivot redshift fixed to $z_0 = 0.62$ \cite{Troxel2018DESy1}, and $\tilde{C}_1 = 5\times10^{-14}\,h^{-2}\mathrm{M}_\odot^{-1}\mathrm{Mpc}^3$ is a normalisation constant. The corresponding angular power spectra of the intrinsic alignment can be expressed as line-of-sight integrals of the three-dimensional power spectra $P_\mathrm{GI}$ and $P_\mathrm{II}$:
\bea
C^{i,j}_{\rm GI}(\ell) &=& \int\dfrac{d\chi}{f_K(\chi)^2} W^i_\gamma(\chi)W^j_N(\chi)P_\mathrm{GI}(k;z) \label{eqn:bnt3_Cgi} \\
C^{i,j}_{\rm IG}(\ell) &=& \int\dfrac{d\chi}{f_K(\chi)^2} W^i_N(\chi)W^j_\gamma(\chi)P_\mathrm{GI}(k;z) \label{eqn:bnt3_Cig} \\
C^{i,j}_{\rm II}(\ell) &=& \int\dfrac{d\chi}{f_K(\chi)^2} W^i_N(\chi)W^j_N(\chi)P_\mathrm{II}(k;z) \label{eqn:bnt3_Cii}
\eea
where $W^i_N(\chi)$ is the number density projection kernel:
\be
W^i_N(\chi) = \frac{H(\chi)}{c} \, n_i(z(\chi)),
\ee
where $H(\chi)$ is the Hubble parameter at $\chi$.

To account for the sky weight induced by the footprint and lensing weights, we employ the \pcl method \citep[PCL,][]{2019namaster} to estimate the angular power spectra of cross-correlations. This will be introduced in Section ~\ref{sec:BNT3_data}.

\subsection{BNT transform}
\label{subsec:IIC}

BNT transform aims to reorganise the data vector $C^{i,j}(\ell)$ to a new $\hat C^{a,b}(\ell)$, where $a$ and $b$ represent the set of linear combinations of the original tomographic redshift bins $i$ and $j$. Throughout the paper, we use the ``hat'' notation $\hat{ }$ for all BNT transformed quantities, and reserve the use of $(i,j)$ for the ordinary, noBNT tomographic bins while using $(a,b)$ for the BNT tomographic bins.

To compute the BNT transform matrix $p^a_i$, we need two normalisation numbers $n_i^{0}$ and $n_i^{1}$ for each tomographic bin $i$:
\bea
n_i^{0} &=& \int\dd\chi \; n_i(\chi) \\
n_i^{1} &=& \int\dd\chi \; \dfrac{n_i(\chi)}{\chi}
\eea
so we can construct the following algebraic equations:
\bea
&\sum_{i=a-2}^a& p^a_{i} n_i^{0} = 0 \\
&\sum_{i=a-2}^a& p^a_{i} n_i^{1} = 0,
\eea
with $p^a_i = 0$ when $i \notin \{a-2, a-1, a\}$, $p_1^1 = p_2^2 = 1$, and $p_2^1 = -1$. The matrix $p^a_i$ is hence a $n_{\rm T}\times n_{\rm T}$ square matrix, where $n_{\rm T}$ is the number of tomographic bins, with nonzero elements only in a diagonal band of a width of 3\gsm{. Here, we show the numerical BNT matrix obtained for the six KiDS-Legacy tomographic bins used in this analysis:}
%(Please see Figure~\ref{fig:BNT_pai_kids}).
\be
\begin{bmatrix}
1.0 & 0.0 & 0.0 & 0.0 & 0.0 & 0.0 \\
-1.0 & 1.0 & 0.0 & 0.0 & 0.0 & 0.0 \\
0.231 & -1.231 & 1.0 & 0.0 & 0.0 & 0.0 \\
0.0 & 1.052 & -2.053 & 1.0 & 0.0 & 0.0 \\
0.0 & 0.0 & 0.305 & -1.304 & 1.0 & 0.0 \\
0.0 & 0.0 & 0.0 & 0.717 & -1.717 & 1.0
\end{bmatrix}. \label{eqn:BNT_pai_kids}
\ee
This example matrix is calculated using the KiDS redshift distribution shown in Figure~\ref{fig:nz_kls} and the so-called ``true background cosmology'' (TBC) used in our previous works \citep[][]{Gu2025BNT1,Gu26BNT2}. \gsm{The KiDS-Legacy dataset and its tomographic redshift distributions are described in detail in Section~\ref{subsec:IIIA}; here we use them only to illustrate the construction of the BNT transform matrix.}

%\begin{figure}[htbp]
%\centering
%\includegraphics[width=0.45\textwidth, trim = 0.0cm 0.0cm 0.0cm 0.0cm, clip]{BNT_Matrix_KiDS.png}
%\caption{
%The BNT transform matrix $p^a_i$. It is calculated with the ``TBC'' cosmology in paper I and II \citep[the ``True Background Cosmology'' in]{Gu2025BNT1,Gu26BNT2} and KiDS Legacy redshift distribution.
%} \label{fig:BNT_pai_kids}
%\end{figure}

\gsm{Following} \citet{Gu2025BNT1}, we apply the BNT transform to the data vector:
\be
\hat C^{a,b}(\ell) = p^a_ip^b_jC_\ell^{i,j}
\label{Cell_BNT}
\ee
and apply $\ell-$cut based on this new data vector. 

\section{Data and Likelihood}
\label{sec:BNT3_data}

\subsection{KiDS-Legacy}
\label{subsec:IIIA}

\begin{figure}[htbp]
\centering
\includegraphics[width=0.47\textwidth, trim = 0.0cm 0.2cm 0.1cm 0.0cm, clip]{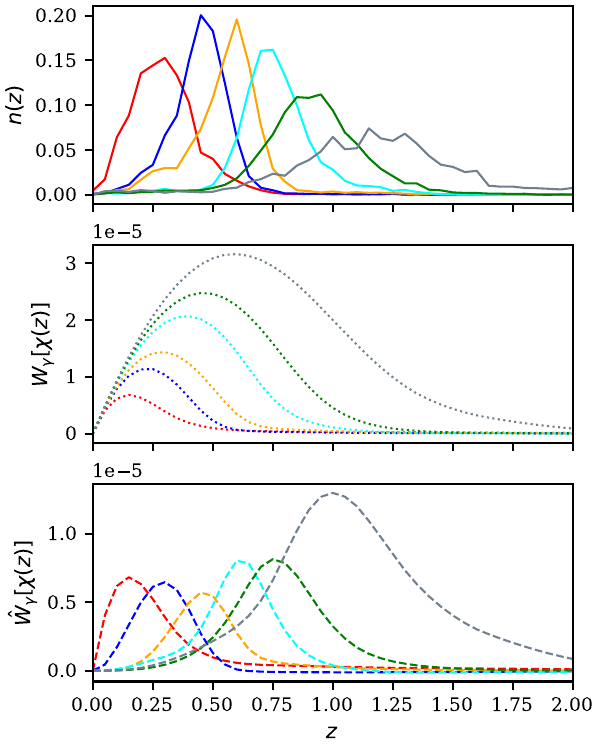}
\caption{Panel 1 of 3: The redshift distribution of the KiDS-Legacy Survey.
Panel 2 of 3: The lensing kernel of the corresponding redshift distribution of panel 1.
Panel 3 of 3: The BNT-transformed lensing kernel from panel 2 with the transformation matrix shown in the Equation~\eqref{eqn:BNT_pai_kids}.
} \label{fig:nz_kls}
\end{figure}

In this work, we make use of the KiDS-Legacy cosmic shear dataset, drawn from the fifth and final data release \cite{Wright24KiDSDR5} of the Kilo Degree Survey (KiDS; \cite{DeJong2013KiDS}). KiDS is a wide-field photometric imaging survey carried out by the European Southern Observatory, combining observations from two facilities: the VLT Survey Telescope (VST; \cite{Enard86VLT,Capaccioli05VST}) and the Visible and Infrared Survey Telescope for Astronomy (VISTA; \cite{Emerson10VISTA,Sutherland15VISTA}). The survey provides measurements of galaxy positions, shapes, and orientations, which form the basis of cosmic shear analyses. Relative to the previous fourth data release, KiDS-1000 \cite{Kuijken19KiDSDR4,Giblin21KiDS1000Catalogue}, which covered $1006\;\mathrm{deg}^2$ before masking ($773.3\;\mathrm{deg}^2$ after masking), the fifth data release expands the survey area to $1347\;\mathrm{deg}^2$ before masking ($967.4\;\mathrm{deg}^2$ after masking). In addition, the photometric redshift limit of the lensing sample is extended from $z=1.2$ to $z=2$, while the effective source number density increases from $7.55\;\mathrm{arcmin}^{-2}$ to $8.79\;\mathrm{arcmin}^{-2}$. These improvements result in a $45.6\%$ increase in the number of source galaxies.

The galaxy samples are constructed using imaging in four optical bands ($u$, $g$, $r$, $i$) from the VST with OmegaCAM \cite{Kuijken2011OmegaCAM}, together with five near-infrared bands ($Z$, $Y$, $J$, $H$, $K_s$) from the VIKING survey (VISTA Kilo-Degree Infrared Galaxy Survey; \cite{Edge13VIKING}).
Compared to the fourth data release, the fifth release introduces an additional set of repeated $i$-band observations reaching greater depth. These deeper $i$-band measurements are specifically designed to improve the robustness of the photometric redshift quality \cite{Wright24KiDSDR5}.
Beyond the wide-field photometric data, the KiDS-Legacy survey also includes the $27\;\mathrm{deg}^2$ KiDZ dataset, which consists of observations of several identified deep spectroscopic survey fields (e.g., \cite{York00SDSS,Blake08WiggleZ,Blake162DFLensingSurvey,DESI1,Drier22GAMADR4}), are used for improving the redshift calibration of the main survey.

In the KiDS-Legacy analysis, photometric redshifts for source galaxies are estimated using the \textsc{BPZ} (Bayesian photo-$z$, \cite{Benitez00BayesianPhotoz}) algorithm. The photometric redshift calibration is carried out using a combination of colour-based self-organising map (SOM; \cite{SOM,Geach12SOM,Wright20SOM}) techniques and spatial clustering methods, with the KiDZ galaxy sample providing accurate spectroscopic redshift information. The extension of the KiDS lensing sample to $z = 2$ allows the source population to be divided into a larger number of tomographic bins than was possible in the KiDS-1000 analysis. This increase in tomographic resolution is particularly important for the application of the BNT formalism, which benefits from (and in practice requires) a sufficiently large number of redshift bins. For this reason, KiDS-Legacy provides a natural and well-matched dataset for testing the BNT-based pipeline developed in \cite{Gu2025BNT1,Gu26BNT2}. Accordingly, the KiDS-Legacy source galaxies are divided into six tomographic bins defined by their \textsc{BPZ} redshift estimates: $0.10 < z_B^1 \leq 0.42 < z_B^2 \leq 0.58 < z_B^3 \leq 0.71 < z_B^4 \leq 0.90 < z_B^5 \leq 1.14 < z_B^6 \leq 2.00$. The inferred redshift distributions, together with the corresponding lensing kernels and their BNT-transformed counterparts, are shown in Figure~\ref{fig:nz_kls}.

The galaxy ellipticities are measured and calibrated by the \textsc{lensfit} dedicated to KiDS DR5. We direct the reader to \citet{Wright24KiDSDR5} for technical details. In this work, we pixelise the ellipticity into a \textsc{HEALPix} scheme \citep[][]{Gorski_2005} to measure the \pcl. The ellipticity in each pixel is given as:

\begin{equation}
    e_{s}(p) = \frac{\sum_{i\in p}w_i e_{s, i}}{\sum_{i\in p}w_i},
\end{equation}
\gsm{where $p$ labels a \textsc{HEALPix} pixel, $i$ labels galaxies within that pixel, $s=1,2$ denotes the two ellipticity components, $e_{s,i}$ is the $s^{\mathrm{th}}$ ellipticity component of galaxy $i$, and $w_i$ is its \textsc{lensfit} weight.} Before pixelisation, the multiplicative and additive bias have been corrected in the ellipticities.

Weight maps are required to measure the \pcl. Here we use the ``SW'' weight map given in \citet{Nicola_2021}:

\begin{equation}
    w(p) = \sum_{i\in p}w_i.
\end{equation}

\subsection{Pseudo-$C_\ell$}
\label{subsec:IIIB}

In this work, we use the \pcl method \citep[][]{Alonso_2018} to practically estimate the angular power spectra. This approach effectively quantifies 2-point functions in harmonic space, offering a direct connection to theoretical predictions. Additionally, it adeptly accounts for the ``mode coupling'' arising from the footprint mask and addresses anisotropic weighting (such as footprint, lensing weight, and partially covered pixels) factors within the footprint. 

Pseudo-$C_{\ell}$ can be applied to correlations between spin-0 (such as galaxy number density) and spin-2 (such as galaxy shapes) fields. In this study, we only consider cosmic shear, so we only present the \pcl formalism for spin-2 $\times$ spin-2. Assume that the sky maps of two fields $u$ and $v$, with corresponding spins $s_u$ and $s_v$, are the products of the underlying whole-sky fields and their respective weights, represented as $w^u(\hat{n})$ and $w^v(\hat{n})$, where $\hat{n}$ signifies a unit vector specifying the angular position. Specifically, for binary masks that delineate only the footprint shape, $w^u(\hat{n})$ equals 1 within the footprint and 0 elsewhere. Throughout the subsequent derivations, we use a ``tilde'' ($\tilde{}$) notation to signify quantities associated with the weighted sky.

The coupled $C_{\ell}$ is measured as:

\begin{equation}
    \tilde{\boldsymbol{C}}_{\ell} ^ {uv} = \frac{1}{2\ell+1}\sum_{m}\tilde{\boldsymbol{a}}_{\ell m}^u\tilde{\boldsymbol{a}}^{v\dagger}_{\ell m},
    \label{eq:cell_def}
\end{equation}
where $\tilde{\boldsymbol{a}}_{\ell m}^u$ is the harmonic coefficient of field $u$ on weighted sky and $\tilde{a}^{u \dagger}_{\ell m}$ is its complex conjugate. We use the bold symbol $\boldsymbol{a}_{\ell m}$ to indicate that a spin-2 field has two components, namely the $E$- and $B$-modes. The bold symbol $\tilde{\boldsymbol{C}}_{\ell}$ indicates the multiple components of \pcl's. In the following derivation, we use the following order $\boldsymbol{C}_{\ell}=\left(C^{EE}_{\ell}, C^{EB}_{\ell}, C^{BE}_{\ell}, C^{BB}_{\ell} \right)$

The coupled $C_{\ell}$ is linked to the underlying power spectra $C_{\ell}$ via:

\begin{equation}
    \tilde{\boldsymbol{C}}_{\ell} ^{uv} = \sum_{\ell^{\prime}} M_{\ell\ell^{\prime}}(w_u, w_v) {\boldsymbol{C}}_{\ell} ^{uv},
\end{equation}
where $M_{\ell\ell^{\prime}}(w_u, w_v)$ is the mode-mixing matrix determined by the weight maps and spins of the two fields. Given the angular power spectrum of the cross-correlation between the weight maps $P_{\ell}\equiv\sum_{m=-\ell}^{\ell}w^u_{\ell m}w^{v *}_{\ell m}$, where $w^u_{\ell m}$ is the harmonic coefficient of the weight map corresponding to field $u$, the forms of the mode-coupling matrix for cosmic shear is given by:

\begin{equation}
\begin{aligned}
M^{22}_{\ell\ell^{\prime}} &= \begin{pmatrix}
M^{+}_{\ell \ell^{\prime}} & 0 & 0 & M^{-}_{\ell \ell^{\prime}}\\
0 & M^{-}_{\ell \ell^{\prime}} & -M^{-}_{\ell \ell^{\prime}} & 0 \\
0 & M^{-}_{\ell \ell^{\prime}} & M^{-}_{\ell \ell^{\prime}} & 0 \\
M^{-}_{\ell \ell^{\prime}} & 0 & 0 & M^{+}_{\ell \ell^{\prime}}
\end{pmatrix}, \mathrm{where} \\
M^{\pm}_{\ell\ell^{\prime}} &= \delta_{\ell \ell^{\prime}}\left(\frac{2\ell^{\prime}+1}{4\pi}\right) \\
&\times\sum_{\ell^{\prime\prime}}P^{uv}_{\ell^{\prime\prime}} 
    \begin{pmatrix}
  \ell^{\prime}\, & \ell^{\prime}\, & \ell^{\prime\prime}\, \\
  2\, & -2\, & 0 \,
 \end{pmatrix}^2 \frac{1\pm(-1)^{\ell+\ell^{\prime}+\ell^{\prime\prime}}}{2}.
\end{aligned}
\label{eq:mode-mix}
\end{equation}

In the formula above, \yanza{each $M$ matrix should be understood as a $N_{\ell}\times N_{\ell}$ block matrices applied on one component of $C_{\ell}$, where $N_{\ell}$ is the number of $\ell$ modes considered in the measurement. Therefore, the expression of $M^{22}_{\ell\ell^{\prime}}$ shows the blocks that are applied on $\boldsymbol{C}_{\ell}\equiv\left(C^{EE}_{\ell}, C^{EB}_{\ell}, C^{BE}_{\ell}, C^{BB}_{\ell} \right)$, respectively. For example, $\tilde{{C}}^{EE}_{\ell}=\sum_{\ell^{\prime}}M^+_{\ell\ell^{\prime}}{{C}}^{EE}_{\ell^{\prime}}+\sum_{\ell^{\prime}}M^-_{\ell\ell^{\prime}}{{C}}^{BB}_{\ell^{\prime}}$. In addition, $\delta_{\ell\ell^{\prime}}$ is the Kronecker delta symbol, $P^{uv}_{\ell}$ is the cross-correlation coupled \pcl of masks corresponding to field $u$ and $v$, and the terms in the brackets with two rows and three columns are the $3-j$ symbol. {Note that only $\ell ''$'s are summed over, the other repeated subscripts $\ell$ and $\ell '$ are free matrix indices and are not summed over.}}

The true underlying power spectra can be calculated by multiplying the coupled $\tilde{\boldsymbol{C}}_{\ell}$ by the inverse mode coupling matrix. In practice, it is more usual to bin the coupled pseudo-$C_{\ell}$ into bandpowers\footnote{Note that the bandpower here refers to binned \pcl, in accordance with \citet{Alonso_2018}. It is not the bandpower used in \citet{Wright2025KiDSLegacy}.}, which will reduce the number of data points without losing information. A bandpower is the weighted $C_{\ell}$ within a range (an $\ell$-bin) of $\ell$. {If we use $q$ to denote the $\ell$-bin index, $\tilde{w}_q^{\ell}$ as the weights at each $\ell$-mode in the bin, and $\vec{\ell}_q$ as the set of $\ell$'s in the $q^{\mathrm{th}}$ bin, the bandpower of coupled pseudo-$C_{\ell}$ in the $q^{\mathrm{th}}$ bin is defined as}:
\begin{equation}
\begin{aligned}
     \tilde{\textbf{\textsf{C}}}_q &= \sum_{\ell \in \Vec{\ell}_q}\tilde{w}_q^{\ell}\tilde{\boldsymbol{C}}_{\ell}   \\
     & = \sum_{\ell \in \Vec{\ell}_q}\sum_{\ell^{\prime}}\tilde{w}_q^{\ell}M_{\ell\ell^{\prime}}{\boldsymbol{C}_{\ell^{\prime}}}.  
\end{aligned}
\label{eq:bp_ps}
\end{equation}
Here we omit the field and spin indices.

To obtain a good estimation of bandpower of the underlying angular power spectra, we introduce the mode-mixing matrix of bandpower by restricting $\ell '$ in the ${q'}^{\mathrm{th}}$ $\ell$-bin:

\begin{equation}
    \boldsymbol{\textsf{M}}_{qq^{\prime}}\equiv \sum_{\ell \in \Vec{\ell}_q}\sum_{\ell^{\prime} \in \Vec{\ell}_{q^{\prime}}}\tilde{w}_q^{\ell}M_{\ell\ell^{\prime}}.
    \label{eq:mode-mix_binned}
\end{equation}

Then the bandpower of coupled pseudo-$C_{\ell}$ can be formally written as:

\begin{equation}
    \tilde{\textbf{\textsf{C}}}_q = \sum_{q^{\prime}}\boldsymbol{\textsf{M}}_{qq^{\prime}}\textbf{\textsf{C}}_{q^{\prime}},
    \label{eq:binned_coupled_cell}
\end{equation}
where $\textbf{\textsf{C}}_{q^{\prime}}$ is the bandpower of the true $\boldsymbol{}_{\ell}$, which can be recovered with the inverse matrix of $\boldsymbol{\textsf{M}}_{qq^{\prime}}$:
\begin{equation}
    \textbf{\textsf{C}}_q = \sum_{q^{\prime}}\left(\boldsymbol{\textsf{M}}\right)^{-1}_{qq^{\prime}}\tilde{\textbf{\textsf{C}}}_{q^{\prime}}.
    \label{eq:bp_true}
\end{equation}
It should be noted that $\textbf{\textsf{C}}_q$ is not simply the true $\boldsymbol{C}_{\ell}$ binned with $\tilde{w}_q^{\ell}$. Comparing Equation~\eqref{eq:bp_true} with Equation~\eqref{eq:bp_ps}, one can connect the bandpower with the unbinned true $\boldsymbol{C}_{\ell}$:

\begin{equation}
    \textbf{\textsf{C}}_q = \sum_{\ell}w_q^{\ell} \boldsymbol{C}_{\ell} = \sum_{q^{\prime}}\left(\boldsymbol{\textsf{M}}\right)^{-1}_{qq^{\prime}} \sum_{\ell \in \vec{\ell}_q}\sum_{\ell^{\prime}}\tilde{w}_q^{\ell}M_{\ell\ell^{\prime}}{\boldsymbol{C}_{\ell^{\prime}}^{uv}},
    \label{eq:binned_true_cell}
\end{equation}
where $w_q^{\ell}\equiv \sum\limits_{q^{\prime}}\left(\boldsymbol{\textsf{M}}\right)^{-1}_{qq^{\prime}} \sum\limits_{\ell \in \Vec{\ell}_{q'}}\tilde{w}_{q'}^{\ell '}M_{\ell^{\prime}\ell}$ is the bandpower window function for the true $C^{uv}_{\ell}$. In practice, we cannot recover $\boldsymbol{C}^{uv}_{\ell}$ from $\textbf{\textsf{C}}^{uv}_q$ since $w_q^{\ell}$ is generally noninvertible. Instead, we compare the measured $\textbf{\textsf{C}}^{uv}_q$ with the bandpower of the theoretical $\boldsymbol{C}^{uv}_{\ell}$ by binning the theoretical $\boldsymbol{C}^{uv}_{\ell}$ with the bandpower window function $w_q^{\ell}$.

In summary, the bandpower of pseudo-$C_{\ell}$ is measured with the following steps:

\begin{enumerate}
    \item directly measure the coupled power spectra $\tilde{\boldsymbol{C}}^{uv}_{\ell}$ with Equation~\eqref{eq:cell_def};
    \item calculate the mode-mixing matrix $M_{\ell\ell'}$  with Equation~\eqref{eq:mode-mix}
    \item bin the mode-mixing matrix with the given binning weight $\tilde{w}_q^{\ell}$ and get $\boldsymbol{\textsf{M}}_{qq^{\prime}}$ with Equation~\eqref{eq:mode-mix_binned};
    \item decouple the bandpower with Equation~\eqref{eq:bp_true} and get the bandpower $\boldsymbol{\textsf{C}}^{uv}_q$ for the true power spectra.
    \item obtain the binning weight for real angular power spectra with Equation~\eqref{eq:binned_true_cell} for the following theoretical fitting.
\end{enumerate}

Another consideration pertains to the smoothing effect introduced by the instrumental beam and the pixelisation window function. Assuming the smoothing is isotropic, it can be characterised by a window function denoted as $b_{\ell}$ in the harmonic space. The overall window function, encompassing all the smoothing effects, results from the multiplication of the individual window functions associated with each smoothing process. Consequently, the measured pseudo-$C_{\ell}$ becomes $\tilde{C}^{uv}_{\ell}\rightarrow b^u_{\ell}b^v_{\ell}\tilde{C}^{uv}_{\ell}$, where $b^u_{\ell}$ represents the overall window function for the field $u$. The PSF size of KiDS DR5 is approximately 1 arcsecond, significantly smaller than the \textsc{HEALPix} pixel size employed in our analysis (1.7 arcminutes corresponding to \texttt{Nside}=2048). Therefore, we will ignore the smoothing effect of the instrumental beam here.

 The autocorrelation \pcl measured following the method above contains autocorrelated noise, namely the shape noise for cosmic shear. For angular power spectra, both are constant with shape noise equal to the shape variance $\sigma_e^2$ divided by galaxy angular number density. For \pcl, we need to calculate the ``mode-coupled'' noise terms, then decouple them.

The coupled shape noise is given by \citet{Nicola_2021}:

\begin{equation}
    \tilde{N}_{\ell} = \Omega_{\mathrm{pix}}\left\langle \sum_{i \in p}w_i^2\sigma_{e,i}^2 \right\rangle_{\mathrm{pix}},
\end{equation}
where $\Omega_{\mathrm{pix}}$ is the angular area of a pixel (in steradian); $\sigma_{e,i}^2\equiv (e_1^2+e_2^2)/2 $ is the variance of galaxy ellipticities; and $\left\langle \cdot \right\rangle_{\mathrm{pix}}$ is the average of pixel values across the whole map. After getting the coupled noise terms, they are decoupled following the same procedure described above, then subtracted from the \pcl. 

For a more comprehensive theoretical treatment of pseudo-$C_{\ell}$ measurements, we direct the reader to \cite{Alonso_2018}. The implementation of pseudo-$C_{\ell}$ measurements is available in the publicly accessible \textsc{NaMaster} package\footnote{\href{https://namaster.readthedocs.io/en/latest/}{https://namaster.readthedocs.io/en/latest/}}, which will be employed in this study. To maintain consistency with existing literature, throughout the remainder of this paper, we continue to use $C_{\ell}$ to denote binned angular power spectra. Here, $\ell$ should be understood as the ``effective $\ell$-mode'' corresponding to each bin. In this work, the BNT transformation will be applied to \pcl measured from the KiDS-Legacy data.

\subsection{Scale cuts}
\label{subsec:IIIC}

\begin{figure*}[htbp]
\centering
\includegraphics[width=0.95\textwidth, trim = 6.5cm 6cm 6cm 6cm, clip]{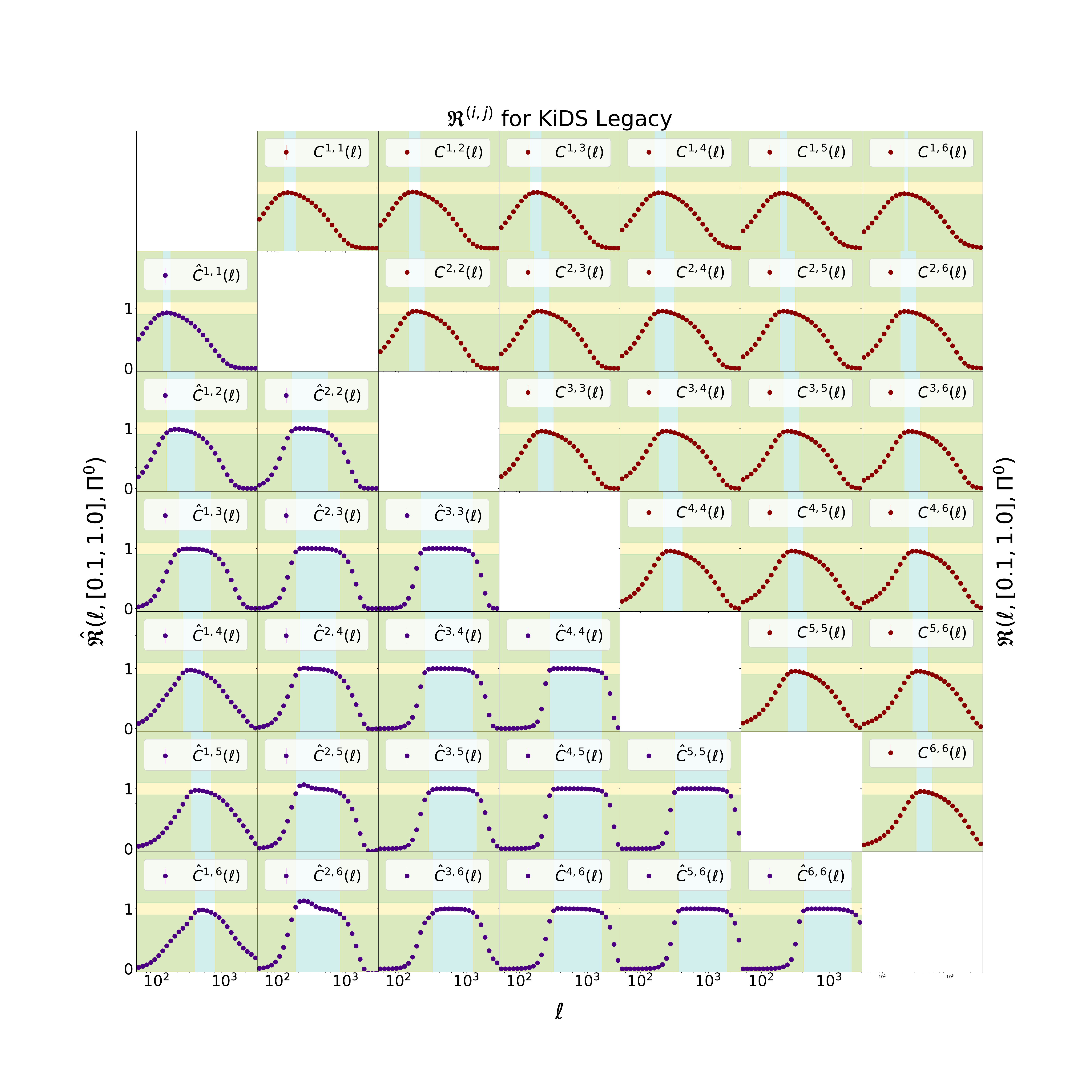}
\caption{Ratio of the data vector with $k_{\rm cut}$ to the full data vector (defined in Equation~\eqref{eqn:bnt3_Rfrak_def}) for the noBNT case $\fR^{(i,j)}(\ell;k_{\rm cuts},\Pi^0)$ (upper right triangle) and the BNT case $\hfR^{(a,b)}(\ell;k_{\rm cuts},\Pi^0)$ (lower left triangle). Each panel shows the ratio for the combination of tomographic bins indicated by $(i,j)$ (for the noBNT case) and $(a,b)$ (for the BNT case). For each tomographic bin, the curves are used to define which cut in $\ell$ corresponds to the constraint that modes $k>k_{\rm cut}^*$ and $k<k_{\rm cut}^\dagger$ should not bias the data vector at a level exceeding $\mathcal{T}_\mathrm{FD}$ for $\ell_{\rm cut}^{\dagger^{(i,j)}}<\ell<\ell_{\rm cut}^{*{(i,j)}}$ for noBNT (resp. $\hat{\ell}_{\rm cut}^{\dagger^{(a,b)}} < \ell<\hat{\ell}_{\rm cut}^{*^{(a,b)}}$ for BNT). All plots are given for $k_\mathrm{cuts} = (0.1\;{\rm Mpc^{-1}},1.0\;{\rm Mpc^{-1}})$, and the horizontal bright band indicates the fractional threshold $\mathcal{T}_\mathrm{FD} = \pm 0.1$. For each tomographic bin combination, the cut in $\ell$ is shown as the boundary of the cyan and green regions.
}
\label{fig:kids_Cl}
\end{figure*}

We adopt the methodology introduced in paper I \cite{Gu2025BNT1} and further developed in paper II \cite{Gu26BNT2}, which provides a framework to identify the $\ell$-range $\ell_{\rm low} < \ell < \ell_{\rm high}$ corresponding to a given physical $k$-band interval $k_{\rm low} < k < k_{\rm high}$ for a specified level of accuracy. Within this $\ell$-interval, the resulting data vector remains accurate to the desired precision. The mapping between $k$-space and $\ell$-space cuts, for both BNT-transformed and conventional (noBNT) data vectors, is quantified using the dimensionless ratios $\hfR$:
\bea
\hfR^{(a,b)} (\ell;k_{\rm cuts},\Pi^0) &\equiv& \frac{\hmC^{a,b}_{[k_\mathrm{low},k_\mathrm{high}]}(\ell;\Pi^0)}{\hmC^{a,b}_{[k_\mathrm{min},k_\mathrm{max}]}(\ell;\Pi^0)} \nonumber\\
{\fR}^{(i,j)}(\ell;k_{\rm cuts},\Pi^0)&\equiv&\frac{{\cal C}^{i,j}_{[k_\mathrm{low},k_\mathrm{high}]}(\ell;\Pi^0)}{{\cal C}^{i,j}_{[k_\mathrm{min},k_\mathrm{max}]}(\ell;\Pi^0)},
\label{eqn:bnt3_Rfrak_def}
\eea
where $k_\mathrm{min} = 0.0025\;\mathrm{Mpc}^{-1}$ and $k_\mathrm{max} = 100\;\mathrm{Mpc}^{-1}$ define the full computational domain in $k$-space. The $\hfR$ ratios quantify the fraction of angular power that remains after applying a $k$-band cut at each $\ell$, comparing the $k$-limited spectrum (numerator) to the full-spectrum case (denominator). As emphasised in paper I, the BNT transform does not require the use of the true cosmological parameters.

To translate a $k$-band cut into the corresponding $\ell$-range, we introduce a tolerance region defined by $\hfR = 1 \pm \tFD$, where $\tFD$ sets the allowed fractional deviation. This parameter can be chosen by the user according to the maximum bias in the angular power spectrum that is acceptable, and is generally determined by the precision required for a given cosmological parameter (e.g., $S_8$) to discriminate between different cosmological models or independent measurements (such as Planck versus weak lensing constraints on $S_8$). In our implementation, a single value of $\tFD$ is applied uniformly to all elements of the data vector. To account for the effects of both $k_\mathrm{cuts}$ and $\tFD$, we employ the continuous Boolean weight function $\mathcal{V}(\ell;k_\mathrm{cuts},\tFD)$:
\be \label{eqn:bnt2_ellcut}
{\mathcal{V}}(\ell;k_\mathrm{cuts},\mathcal{T}_\mathrm{FD}) = \begin{cases}
1, |\fR (\ell;k_{\rm cuts},\Pi^0)-1| \leq \mathcal{T}_\mathrm{FD} \\
0, |\fR (\ell;k_{\rm cuts},\Pi^0)-1| > \mathcal{T}_\mathrm{FD},
\end{cases}
\ee
where $\fR(\ell;k_{\rm cuts},\Pi^0)$ represents the dimensionless ratio of the angular power spectra computed from the full tomographic data vector. This definition applies equally to both the BNT-transformed and conventional (noBNT) data vectors. In the likelihood evaluation, the Boolean weight function is applied as follows:
\bea
\Delta {\cal D}^* &\equiv& \mathcal{V}(\ell;k_\mathrm{cuts},\mathcal{T}_\mathrm{FD})\left({\cal C}(\Pi)-{\cal C}(\Pi^0)\right) \nonumber\\
\mathbb{C^*} &\equiv& \mathcal{V}(\ell;k_\mathrm{cuts},\mathcal{T}_\mathrm{FD}) \cdot \mathbb{C} \cdot \mathcal{V}(\ell;k_\mathrm{cuts},\mathcal{T}_\mathrm{FD})^{^T} \nonumber \\ %\label{star_eqn}
\chi^2(\Pi;k_\mathrm{cuts},\mathcal{T}_\mathrm{FD})&=&\Delta {{\cal D}^*}^T \cdot (\mathbb{C^*})^{-1} \cdot \Delta {\cal D}^*,\label{eqn:bnt3_chi2kFD}
\eea
where ${\cal C}(\Pi)$ denotes the full tomographic data vector. Its structure follows a systematic ordering of all tomographic cross- and autopower spectra:
\be \label{eqn:tomo}
{\cal C}_{[k_1,k_2]}(\Pi) = \begin{pmatrix}
    \begin{array}{cc}
        \begin{bmatrix}
       \tilde C^{1,1}_{[k_1,k_2]}(\ell;\Pi)  \\
       \end{bmatrix} {\rm all}~\ell's \\
       \begin{array}{c}
       \end{array}\\
       \begin{bmatrix}
        \tilde C^{1,2}_{[k_1,k_2]}(\ell;\Pi) \\
       \end{bmatrix} {\rm all}~\ell's \\
       \begin{array}{c}
       \end{array}\\
       \begin{bmatrix}
       \tilde C^{1,3}_{[k_1,k_2]}(\ell;\Pi) \\
       \end{bmatrix} {\rm all}~\ell's \\
        ...& \\
       \begin{array}{c}
       \end{array}\\
       \begin{bmatrix}
       \tilde C^{5,6}_{[k_1,k_2]}(\ell;\Pi) \\
       \end{bmatrix} {\rm all}~\ell's \\
       \begin{array}{c}
       \end{array}\\
       \begin{bmatrix}
       \tilde C^{6,6}_{[k_1,k_2]}(\ell;\Pi) \\
       \end{bmatrix} {\rm all}~\ell's \\
    \end{array}
\end{pmatrix}.
\ee
In this construction, the $\mathcal{V}$ weight ensures that only the $\ell$-modes consistent with the chosen $k$-band cuts and fractional tolerance $\tFD$ contribute to both the residual $\Delta {\cal D}^*$ and the corresponding covariance matrix $\mathbb{C}^*$ in each case. This formulation applies equally to the BNT-transformed and conventional (noBNT) data vectors, maintaining the integrity of the likelihood evaluation under the $k$-cut scheme.
The distinction between $\tilde C^{i,j}(\ell)$ and $\hat{\tilde C}^{a,b}(\ell)$ lies in the labelling of tomographic bins, with $i,j$ referring to the conventional (noBNT) bins and $a,b$ to the BNT-transformed bins. 
%Correspondingly, $\mathbb{C}$ denotes the covariance matrix associated with each case. 
Figure~\ref{fig:kids_Cl} presents the power spectrum ratios for both BNT and noBNT, computed under a specific $k$-band cut for all tomographic bin combinations in the KiDS-Legacy dataset.

\subsection{Covariance matrices}

The covariance matrix of \pcl is modelled as a summation of a Gaussian covariance, connected non-Gaussian covariance, and supersample covariance (SSC): 
\be
\mathbb{C}_{pC_\ell} = \mathbb{C}^{G}_{pC_\ell} + \mathbb{C}^{nG}_{C_\ell} + \mathbb{C}^{SSC}_{C_\ell} .
\ee
\yanza{The first term $\mathbb{C}^{G}_{pC_\ell}$ is the dominant ``disconnected'' covariance matrix corresponding to Gaussian fields, including physical Gaussian fluctuations and Gaussian noise:}

\begin{equation}\mathbb{C}^{G}\left(C_{\ell_1}^{u v}, C_{\ell_2}^{w z}\right)=\delta_{\ell_1 \ell_2} \frac{C_{\ell_1}^{u w} C_{\ell_2}^{v z}+C_{\ell_1}^{u z} C_{\ell_2}^{v w}}{(2 \ell_1+1)}
\label{eq:covg}
.\end{equation}
\yanza{This is the covariance matrix for an all-sky, uniformly weighted measurement. Sky weights or masks, and binning introduce nonzero coupling between different $\ell$ and components, as well as enlarge the variance. To account for this, we used the method given by \cite{Garc_a_Garc_a_2019}. For simplicity, we present the format for spin-0, unbinned case. The cosmic shear bandpower case follows the similar idea, with more complicated mathematics. In practice, they are computed following the method given by \citet{Garc_a_Garc_a_2019} with narrow kernel approximation \citep[NKA,][]{Nicola_2021}. We direct the reader to \cite{Garc_a_Garc_a_2019} for the details.
When considering sky masks, the disconnected Gaussian covariance is:}

\begin{equation}
\begin{aligned}
\mathbb{C}^{G}_{pC_\ell}\left(C_{\ell_1}^{u v}, C_{\ell_2}^{w z}\right) &=\frac{1}{2 \ell_1+1}\left[C_{(\ell_1}^{uw} C_{\left.\ell_2\right)}^{vz} M_{\ell_1 \ell_2}\left(w_u w_w, w_v w_z\right)\right.\\
& \left.+C_{(\ell_1}^{uz} C_{\left.\ell_2\right)}^{vw} M_{\ell_1 \ell_2}\left(w_u w_z, w_v w_w\right)\right],
\end{aligned}
\end{equation}
\yanza{where}

\begin{equation}
    C_{(\ell_1}^{uw} C_{\left.\ell_2\right)}^{vz} \equiv \frac{1}{2} \left( C_{\ell_1}^{uw} C_{\ell_2}^{vz} + C_{\ell_2}^{uw} C_{\ell_1}^{vz} \right),
    \label{eq:cell_cov}
\end{equation}
\yanza{and $M_{\ell_1 \ell_2}\left(w_u w_w, w_v w_z\right)$ is the mode-mixing matrix of the maps $w_u w_w$ (the multiplication of the mask corresponding to $u$ and $w$ fields) and $w_v w_z$ that is defined in Equation~\eqref{eq:mode-mix}.}

The calculation is conducted with the \textsc{NaMaster} package, given the lensing weight maps. The angular power spectra $C_{\ell}$ to calculate the Gaussian covariance are obtained either from the theoretically calculated data vector using \textsc{pyccl} \citep[][]{Pyccl} or from the observed KiDS-Legacy data vector. For posterior samples obtained using differently constructed covariance matrices, we adopt the following abbreviations: TDC denotes the theoretical-data-vector-based covariance, while ODC denotes the observational-data-vector-based covariance. In all cases, the likelihoods are evaluated using the same KiDS-Legacy observational data vector.

In this work, we only consider the leading term of the connected non-Gaussian covariance and SSC covariance from the one-halo component of the trispectra. Both terms are obtained from the \textsc{OneCovariance} package \citep{OneCov} with the survey geometry from standard KiDS-Legacy footprint described in Section~\ref{subsec:IIIA} and in \citet{Wright2025KiDSLegacy}. The cosmological parameters that we use to calculate theoretical angular power spectra for the Gaussian term, as well as the trispectra for non-Gaussian and SSC covariances, are adopted from KiDS-Legacy cosmic shear cosmology \citep[][]{Wright2025KiDSLegacy}.

\subsection{Analysis setup}
\label{subsec:IIIE}

\begin{table}[t]
\centering
\caption{\label{tbl:prior}%
The prior of all the parameters used in our sampling. $\mathcal{U}(a,b)$ is a flat prior  between $a$ and $b$, and $\mathcal{N}(c,d)$ is a Gaussian prior centred at $c$ with standard deviation $d$. The definition of nuisance parameters below and the correlation matrix of the redshift uncertainty $\delta_{z,i}$ is given in \citet{Wright2025KiDSLegacy}.}
\rowcolors{2}{white}{lightgray!20}
\begin{tabular}{|l|c|c|}
\hline
\rowcolor{cyan!10} % Light cyan background for the top row
\ \textrm{Parameters} \ &
\ \textrm{Fiducial Value} \ & 
\textrm{Priors} \\
\hline
\ $S_8$ & $0.7970$ & $\mathcal{U}(0.5,1.0)$\\
\ $\Omega_c h^2$ & $0.1530$ & $\mathcal{U}(0.051,0.255)$\\
\ $\Omega_b h^2$ & $0.0225$ & $\mathcal{U}(0.019,0.026)$\\
\ $n_s$ & $0.9690$ & $\mathcal{U}(0.83,1.11)$\\
\ $h$ & $0.6898$ & $\mathcal{U}(0.63,0.83)$\\
\ $T_\mathrm{AGN}$ & $7.8000$ & $\mathcal{U}(7.3,8.3)$\\
\ $A_\mathrm{IA}$ & $0.0$ & $\mathcal{U}(-0.6,0.6)$\\
\hline
\ $m_{1}$ & $-0.023$ & $\mathcal{N}(-0.023,0.0060)$ \\
\ $m_{2}$ & $-0.016$ & $\mathcal{N}(-0.016,0.0060)$ \\
\ $m_{3}$ & $-0.011$ & $\mathcal{N}(-0.011,0.0070)$ \\
\ $m_{4}$ & $0.0200$ & $\mathcal{N}(0.0200,0.0070)$ \\
\ $m_{5}$ & $0.0300$ & $\mathcal{N}(0.0300,0.0080)$ \\
\ $m_{6}$ & $0.0450$ & $\mathcal{N}(0.0450,0.0090)$ \\
\hline
\ $\delta_{z,1}$ & $-0.026$ & $\mathcal{N}(-0.026,0.0102)$ \\
\ $\delta_{z,2}$ & $0.0140$ & $\mathcal{N}(0.0140,0.0100)$ \\
\ $\delta_{z,3}$ & $-0.002$ & $\mathcal{N}(-0.002,0.0102)$ \\
\ $\delta_{z,4}$ & $0.0080$ & $\mathcal{N}(0.0080,0.0100)$ \\
\ $\delta_{z,5}$ & $-0.011$ & $\mathcal{N}(-0.011,0.0102)$ \\
\ $\delta_{z,6}$ & $0.0540$ & $\mathcal{N}(0.0540,0.0108)$ \\
%\hline
%\ $\sigma_8$ & $0.8256$ & \ Derived Parameter \ \\
%\ $S_8$ & $0.8124$ & \ Derived Parameter \ \\
\hline
\end{tabular}
%\end{ruledtabular}
\end{table}

%\begin{table}[t]
%\centering
%\caption{\label{tbl:redshift_cov}%
%The correlation matrix $D^\mathrm{CC}_{ij}$ of redshift uncertainties used in our sampling.}
%\rowcolors{2}{white}{lightgray!20}
%\begin{tabular}{|c|c|}
%\hline
%\rowcolor{cyan!10} % Light cyan background for the top row
%\ \textrm{Parameters} \ &
%\ \textrm{Value}\ \\
%\hline
%$D^\mathrm{CC}_{12}$ & -0.09 \\
%$D^\mathrm{CC}_{13}$ & 0.05 \\
%$D^\mathrm{CC}_{23}$ & 0.19 \\
%$D^\mathrm{CC}_{14}$ & 0.02 \\
%$D^\mathrm{CC}_{24}$ & -0.19 \\
%$D^\mathrm{CC}_{34}$ & -0.35 \\
%$D^\mathrm{CC}_{15}$ & 0.00 \\
%$D^\mathrm{CC}_{25}$ & -0.11 \\
%$D^\mathrm{CC}_{35}$ & -0.20 \\
%$D^\mathrm{CC}_{45}$ & 0.14 \\
%$D^\mathrm{CC}_{16}$ & -0.00 \\
%$D^\mathrm{CC}_{26}$ & 0.07 \\
%$D^\mathrm{CC}_{36}$ & -0.22 \\
%$D^\mathrm{CC}_{46}$ & 0.28 \\
%$D^\mathrm{CC}_{56}$ & -0.02 \\
%\hline
%\end{tabular}
%\end{table}

The analysis setup for this work combines the methodology from our paper I \cite{Gu2025BNT1} with the fiducial analysis of KiDS-Legacy \cite{Wright2025KiDSLegacy}. We perform a series of Bayesian posterior estimations to quantify the impact of intrinsic alignment, using 
both a theoretical mock data vector based on a ``true background cosmology'' (TBC) and 
the observational pseudo-$C_\ell$ data vector derived from the KiDS-Legacy shape catalogue. To recapitulate the fiducial cosmological results of the KiDS-Legacy survey, we adopt parameter priors (Please see Table~\ref{tbl:prior} and the Figure 3 of \citet{Wright2025KiDSLegacy} for the $D^\mathrm{CC}$ redshift error correlation matrix) almost identical to those used in the KiDS-Legacy fiducial analysis \cite{Wright2025KiDSLegacy}.

In our main analysis, we use 30 logarithmically spaced $\ell$-bins, with bin centres ranging from $\ell_\mathrm{min} = 50$ to $\ell_\mathrm{max} = 3000$. It is worth noting that the fiducial bandpower analysis, which is different from pseudo-$C_\ell$, of KiDS-Legacy used $\ell_\mathrm{min} = 100$ to $\ell_\mathrm{max} = 1500$. Therefore, we also use our data vector with the corresponding $\ell$-cut to check our consistency with \citet{Wright2025KiDSLegacy}. All theoretical computation in the posterior sampling process, except for the BNT transform and pseudo-$C_\ell$-related computations, is computed with the core cosmology library (\textsc{CCL}) \cite{Pyccl}, a publicly available and standardised library for calculating cosmological observables. \textsc{CCL} is developed and maintained by the LSST Dark Energy Science Collaboration (DESC) \cite{DESC2018DESC}. We sample our parameter posteriors' likelihood with the \textsc{Nautilus} package \cite{Nautilus}, a highly efficient importance-nested sampling toolkit designed for Bayesian posterior and evidence estimation.

\section{Results}
\label{sec:BNT3_results}

\subsection{Consistency check and effect on $\ell_\mathrm{min}$}

\begin{figure*}
\centering
\includegraphics[width=0.90\textwidth, trim = 0.25cm 0.2cm 0.2cm 0.25cm, clip]{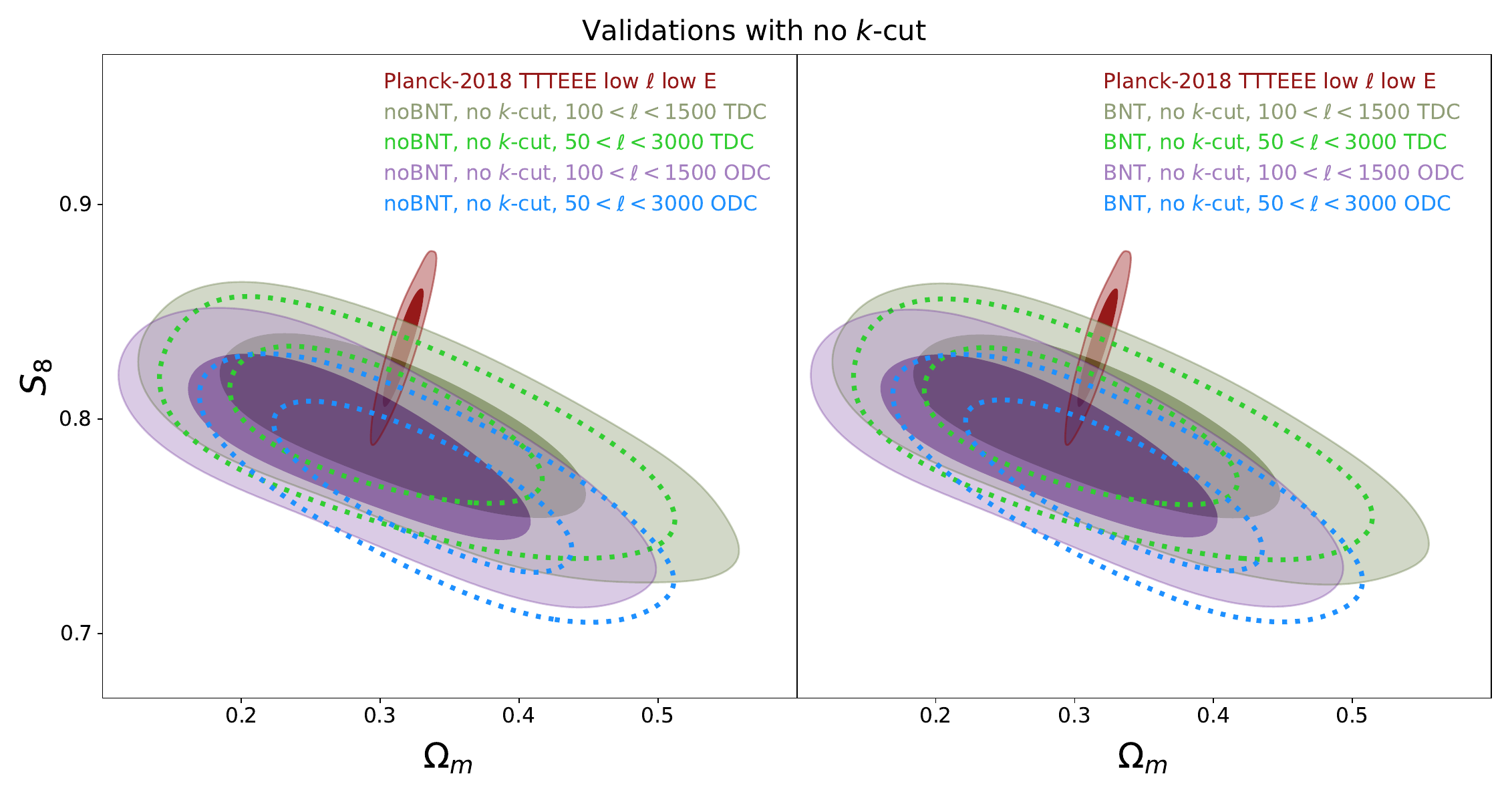}
% trim: Left Bottom Right Top
\caption{Constraints in the $(\Omega_m,S_8)$ plane for the noBNT (left panel) and BNT (right panel) from the KiDS-Legacy pseudo-$C_\ell$ validation posteriors.
For both noBNT and BNT, we show the constraint with four cases: the KiDS-Legacy like $100 < \ell < 1500$ $\ell$-cut (filled contours), and the $50 < \ell < 3000$ $\ell$-cut (dotted contours), both inferred with either the theoretical data-vector-based covariance matrix (TDC) setup (green contours) or the observational data-vector based covariance matrix (ODC) setup (blue contours). The noBNT and BNT contours are expected to reproduce each other when no $k$-cut is applied.
The red filled contour is the TTTEEE-low$\ell$-lowE cosmological posterior coming from \citet{Planck2020Results}.
}
\label{fig:nocut_validation}
\end{figure*}

We first verify that the cosmological constraints derived from our KiDS-Legacy pseudo-$C_\ell$ analysis are consistent with the fiducial KiDS-Legacy results reported in \citet{Wright2025KiDSLegacy}. %Given the close similarity between pseudo-$C_\ell$ and band powers, we adopt $S_8 = 0.797$ from the posterior mean of the KiDS-Legacy Band Power, as shown in Table 4 of \citet{Wright2025KiDSLegacy}. 
Figure~\ref{fig:nocut_validation} demonstrates that our validation posterior recovers $S_8 = 0.787^{+0.031}_{-0.023}$ for the ODC result and $S_8 = 0.795^{+0.030}_{-0.025}$ for the TDC result, in good agreement with the KiDS-Legacy bandpower result of $S_8 = 0.797^{+0.023}_{-0.024}$ for the $\ell$-range $100 < \ell < 1500$, for both noBNT and BNT cases, as they are expected to be identical to each other when there is no physical scale cut applied.

We also note that, in ODC contours when the additional $\ell$-ranges $50 < \ell < 100$ and $1500 < \ell < 3000$ are included in the ODC posteriors, the $S_8$ constraint shifts downward to $S_8 = 0.768^{+0.023}_{-0.023}$. %To identify which $\ell$-range drives this bias, Figure~\ref{fig:nocut_validation} also shows results for the cut $100 < \ell < 3000$, which extends to smaller scales only. The $S_8$ constraint in this case only has a small shift compared to the range without $1500 < \ell < 3000$, indicating that the low-$S_8$ shift in the ODC contours originates primarily from the $50 < \ell < 100$ modes. 
Although this effect is \gsm{modest compared with the posterior width,} it suggests a slight preference for lower $S_8$ values from the extended scales. This trend does not appear in the TDC posteriors. When we include the same additional $\ell$-ranges, $50 < \ell < 100$ and $1500 < \ell < 3000$, the $S_8$ constraint remains almost unchanged, with $S_8 = 0.794^{+0.022}_{-0.022}$. This indicates that the shift in the ODC posteriors does not originate solely from the data vector, but is likely related to the covariance matrix.

\subsection{Constraint with truncations in $k$-space}

\begin{figure*}
\centering
\includegraphics[width=0.90\textwidth, trim = 0.25cm 0.2cm 0.2cm 0.5cm, clip]{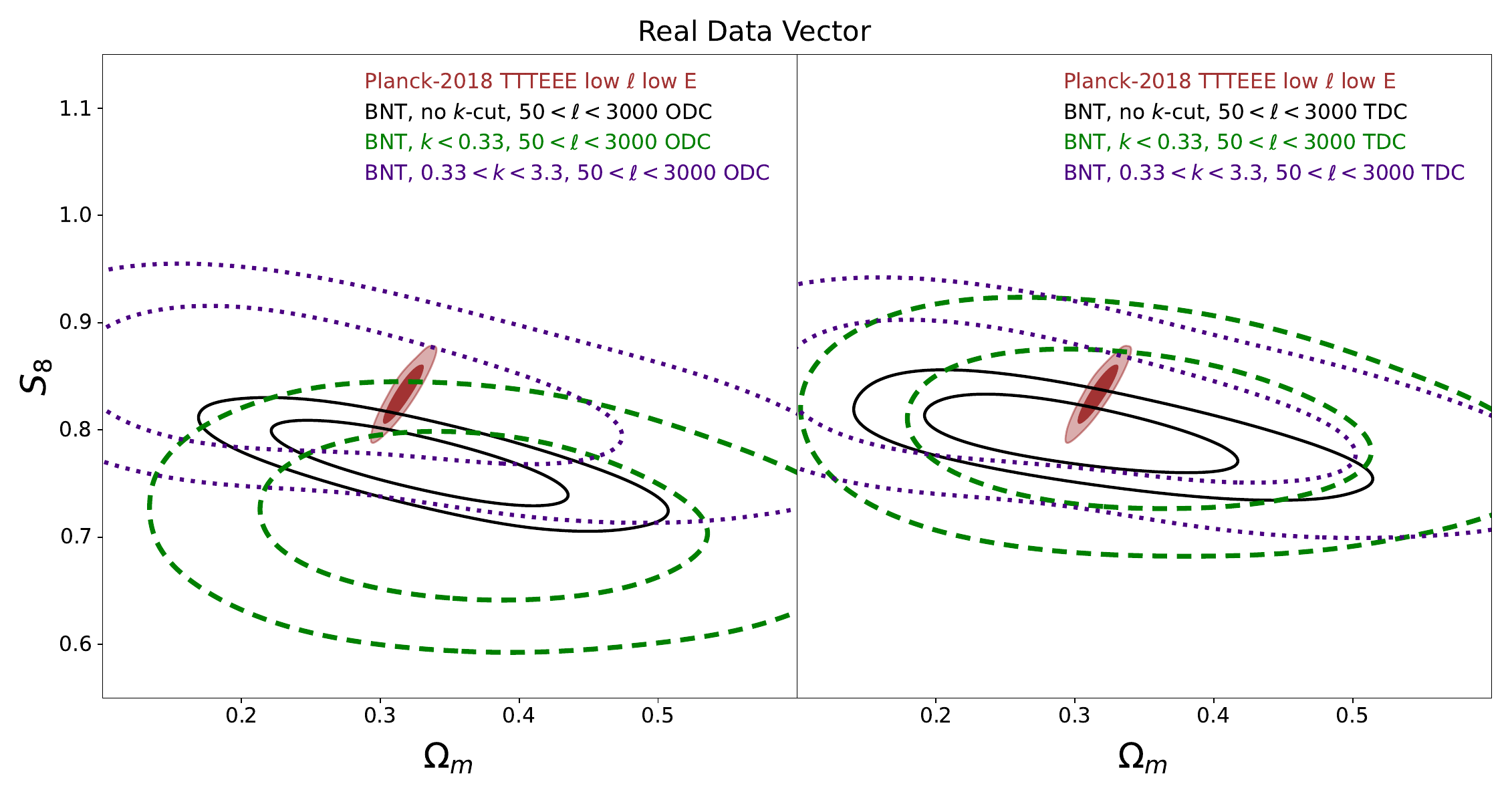}
% trim: Left Bottom Right Top
\caption{Constraints in the $(\Omega_m,S_8)$ plane for the ODC-BNT (left panel) and TDC-BNT (right panel) from the real KiDS-Legacy pseudo-$C_\ell$ posteriors. We show the case without a $k$-cut (black solid) and two contours with different $k$-space scale selections: $k < 0.33\;\mathrm{Mpc}^{-1}$ (green dashed), $k \in (0.33,3.3)\;\mathrm{Mpc}^{-1}$ (indigo dotted). 
For all contours here, we use the fractional difference threshold of $\tFD = 0.1$. The red filled contour is the TTTEEE-low$\ell$-lowE cosmological posterior coming from \citet{Planck2020Results}.}
\label{fig:kcut}
\end{figure*}

In \citet{Gu2025BNT1}, we showed that the exclusion of the small-scale signal through the BNT transform can effectively remove the astrophysical biases while preserving the constraining power on information of cosmological structure formation history. Using the real data, we make the same test and have the following $S_8$ constraint with three $k_\mathrm{high}$ values: $k < 0.1\;\mathrm{Mpc}^{-1}$ (ODC: $S_8 = 0.664^{+0.095}_{-0.150}$; TDC: $S_8 = 0.719^{+0.131}_{-0.159}$), $k < 0.33\;\mathrm{Mpc}^{-1}$ (ODC: $S_8 = 0.717^{+0.047}_{-0.046}$; TDC: $S_8 = 0.798^{+0.045}_{-0.045}$), and $k < 1.0\;\mathrm{Mpc}^{-1}$ (ODC: $S_8 = 0.756^{+0.032}_{-0.032}$; TDC: $S_8 = 0.810^{+0.032}_{-0.030}$)\footnote{Please note that the contour plots are smoothed, whereas the median and error bar values are not.}
, where we show both ODC and TDC contours of $k < 0.33\;\mathrm{Mpc}^{-1}$ in Figure~\ref{fig:kcut}.

\gsm{We see that constraints from different scale cuts differ from one another by less than $1.5\sigma$. We therefore find no large shift due to the removal of nonlinear scales at the current precision of KiDS-Legacy.} However, another interesting pattern is that the larger scales prefer a lower value of $S_8$, particularly in the ODC-based posterior sampling results. For the $k < 0.33~\mathrm{Mpc}^{-1}$ cut, the ODC-based $S_8$ posterior deviates from Planck at the $2.32\sigma$ level. In contrast, the corresponding TDC-based constraint reduces the tension with Planck to $0.76\sigma$. This suppression at the large-scale might originate from the case where the data vector fluctuates downward at large-scales, and the ODC covariance will also fluctuate downward at the large scale. Therefore, the resulting posterior will put more weight on those values than it supposed to be, and hence the posterior would prefer an even lower $S_8$ than in the TDC case.

In \citet{Gu26BNT2}, we also demonstrated that comparing cosmological constraints obtained from different cuts in the $k$-band provides a critical diagnostic to identify potential modelling biases in the posterior sampling process. For the real data, similarly, we have the following $S_8$ constraint with three $k_\mathrm{cuts}$ values: $k \in (0.033,0.33)\;\mathrm{Mpc}^{-1}$ (ODC: $S_8 = 0.749^{+0.056}_{-0.056}$; TDC: $S_8 = 0.806^{+0.051}_{-0.051}$), $k \in (0.1,1.0)\;\mathrm{Mpc}^{-1}$ (ODC: $S_8 = 0.790^{+0.057}_{-0.049}$; TDC: $S_8 = 0.806^{+0.065}_{-0.047}$), and $k \in (0.33,3.3)\;\mathrm{Mpc}^{-1}$ (ODC: $S_8 = 0.833^{+0.044}_{-0.044}$; TDC: $S_8 = 0.819^{+0.044}_{-0.043}$), where we show both ODC and TDC contours of $k \in (0.33,3.3)\;\mathrm{Mpc}^{-1}$ in Figure~\ref{fig:kcut} as well, showing the largest discrepancy to the $k < 0.33\;\mathrm{Mpc}^{-1}$ in the ODC result.

We find that, for the TDC contours, the $S_8$ constraints obtained from different $k$-bands are highly consistent with each other. This indicates that the cosmological \gsm{inferences} using the theoretical data-vector-based covariance matrix is \gsm{internally self-consistent.} However, we observe a $S_8$ trend in the ODC contours--a $1.24\sigma$ deviation between the constraints from the largest-scale bin, $k \in (0.033,0.33)$, and the smallest-scale bin, $k \in (0.33, 3.3)$. A similar, though weaker, trend, although not shown in figures, is also present in the noBNT contours with $\tFD = 0.1$, yielding $S_8 = 0.774 \pm 0.045$ for $k \in (0.033,0.33)$, $S_8 = 0.811 \pm 0.036$ for $k \in (0.1, 1.0)$, and $S_8 = 0.828 \pm 0.037$ for $k \in (0.33, 3.3)$. This suggests that the observed scale-dependent trend does not originate from the BNT transformation of the data vector or the BNT transformation of the observational data-vector-based covariance matrix.

\subsection{Cosmological pipeline implications}

\begin{figure*}
\centering
\includegraphics[width=0.90\textwidth, trim = 0.25cm 0.2cm 0.2cm 0.5cm, clip]{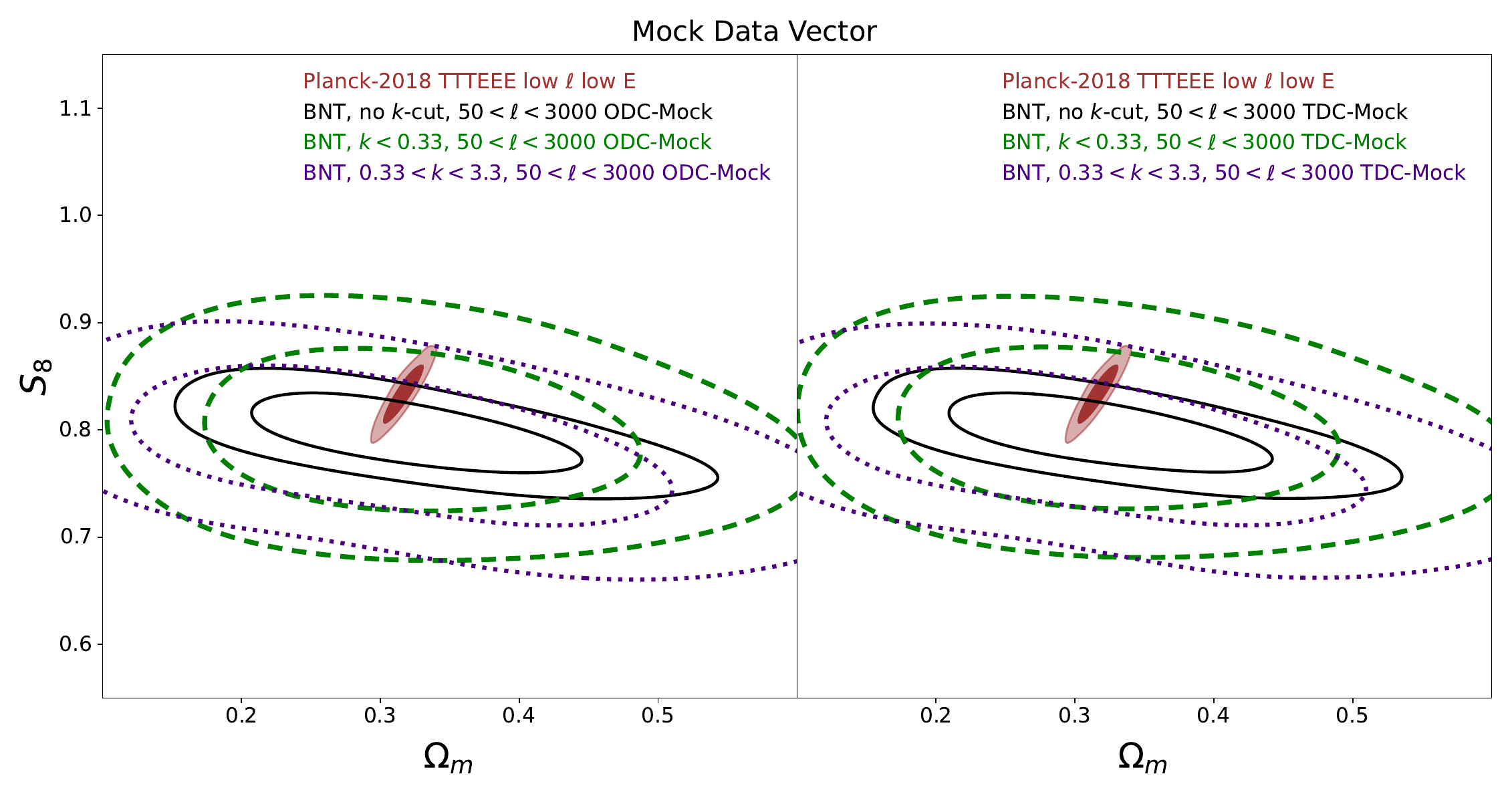}
% trim: Left Bottom Right Top
\caption{Same as Figure~\ref{fig:kcut}, but with theoretical data vector as the mock observation instead.
}
\label{fig:mock}
\end{figure*}

The large-scale bias in the likelihood based on the observational-data-vector covariance matrix does not appear to be a direct consequence of the covariance matrix when tested on a noiseless mock data vector. When the theoretical prediction is used as the mock data vector, the likelihoods obtained with the theoretical-data-vector covariance matrix and the observational-data-vector covariance matrix produce almost the same posterior results. In Figure~\ref{fig:mock}, we show this likelihood test for the mock observable, using the two scale cuts $k \leq 0.33~\mathrm{Mpc}^{-1}$ and $k \in (0.33,3.3)~\mathrm{Mpc}^{-1}$. This pair exhibits the largest mutual deviation in the likelihoods based on the observational data vector with the observational-data-vector covariance matrix. Since this mock data vector does not include noise, this test should be interpreted only as a consistency check of the covariance treatment, rather than a one-to-one reproduction of the data likelihood. Nevertheless, the similarity between the ODC and TDC mock posteriors suggests that the covariance matrix alone is unlikely to be the sole origin of the large-scale deviation seen in the ODC posterior likelihoods with the KiDS-Legacy pseudo-$C_\ell$ data vector.

\begin{figure*}[htbp]
\centering
\includegraphics[width=0.44\textwidth, trim = 0.05cm 0.1cm 0.1cm 0.0cm, clip]{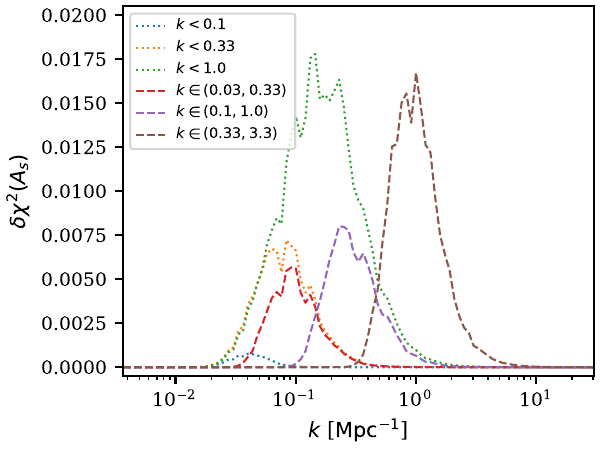}
\includegraphics[width=0.52\textwidth, trim = 0.0cm 0.8cm 0.1cm 0.0cm, clip]{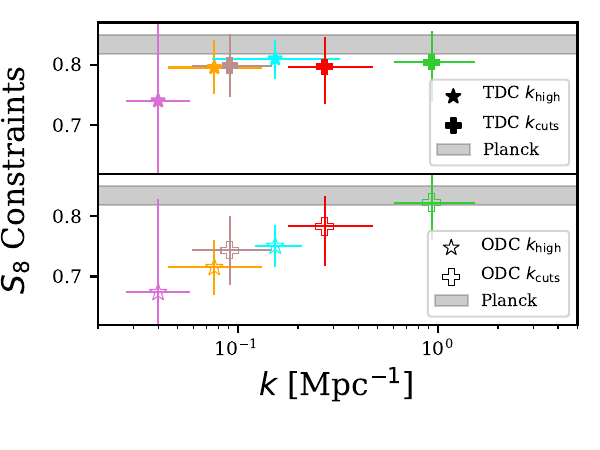}
\caption{Left: the contribution to the total $A_s$ constraining power from each scale to the total $\chi^2$ for different BNT-based $k$-cut choices with $\tFD = 0.1$. 
Right: the constraint of $S_8$ of each $k$-cut choices with TDC (upper half) and ODC (lower half) setups. The $S_8$ error bar of each data point comes from the weighted $16, 50, \mathrm{and}~84\%$ percentage of the $1$-dimensional posterior on $S_8$. Their error bar on $k$-axis is computed from the $16, 50, \mathrm{and}~84\%$ percentage by treating $\delta\chi^2(k)$ as a probability distribution function. The star shape data points indicate the $k_\mathrm{high}$ truncation with only the upper limit on the $k$-range selection, while the cross shape data points indicate $k_\mathrm{cuts}$ truncation that have both the upper and the lower limits on the $k$-range selection. The grey shaded region indicates the $S_8$ constraint coming from the Planck TTTEEE-low$\ell$-lowE result.
} \label{fig:BNT3_deltachi2}
\end{figure*}

In the Appendix of \citet{Gu2025BNT1}, we proposed a method to quantify the scale-dependent contribution of the data vector to each modelling parameter for a given survey specification. For KiDS-Legacy, and for all scale cuts defined by either $k_\mathrm{high}$ or $k_\mathrm{cuts}$, we compute the corresponding scale-wise contributions relative to the total $\chi^2$, evaluated by perturbing $A_s$ by a minimal amount under the fiducial KiDS-Legacy setup. This quantity should be interpreted as a normalised, scale-dependent contribution to the total $\chi^2$, rather than as a direct measure of Fisher information or parameter uncertainty. This procedure yields a scale-dependent distribution of the sensitivity to $A_s$ across $k$-modes, which is shown in the left panel of Figure~\ref{fig:BNT3_deltachi2}. We then estimate the $1\sigma$ interval of this distribution to characterise the range of scales from which each $k$-cut chain receives the dominant contribution.
 
Using the posterior distribution of $S_8$ obtained from the likelihood analysis, we further present the ODC-BNT (empty markers) and TDC-BNT (filled markers) $S_8$ constraints associated with the $k_\mathrm{high}$- and $k_\mathrm{cuts}$-based scale selections in the right panel of Figure~\ref{fig:BNT3_deltachi2}. We find that the TDC $S_8$ constraints are mostly consistent with each other, although the largest-scale cut still shows some suppression. In contrast, the ODC $S_8$ constraints exhibit a much more obvious correlated trend in which larger physical scales, corresponding to lower $k$, prefer lower values of $S_8$. This behaviour is consistently observed across all data points, each corresponding to a posterior analysis with a specific scale cut. The largest difference occurs between the scale ranges $k < 0.33\;\mathrm{Mpc}^{-1}$ (orange stars) and $k \in (0.33,3.3)\;\mathrm{Mpc}^{-1}$ (lime green crosses), reaching a maximum \gsm{difference of $1.80\sigma$ between the posteriors. The coherent trend across the scale selections is absent in the mock tests, suggesting} that the large-scale preference for lower $S_8$ is not a generic consequence of the covariance prescription or the data vector alone. Instead, it may reflect some specific interplays between the KiDS-Legacy pseudo-$C_\ell$ data vector and the observational-data-vector-based Gaussian covariance matrix.

\section{Discussion and Conclusions}
\label{sec:BNT3_conclusion}

Scale cuts in weak lensing data, whether applied in configuration space or harmonic space, are a critical component of analyses aiming to address cosmological parameter tensions, such as the $S_8$ discrepancy. However, because of the two-dimensional projection nature of conventional weak gravitational lensing, scale cuts are imperfect and invariably retain some contribution from unwanted scales.

The BNT transform \cite{Bernardeau2014BNT} is a method that helps suppress information from these unwanted scales while preserving, as much as possible, the relevant information in both directions. Building on this approach, our previous works \cite{Gu2025BNT1, Gu26BNT2} developed a methodological framework for testing theoretical modelling in the posterior sampling process and deriving cosmological constraints. In the present study, we apply this BNT-based analysis pipeline to the KiDS-Legacy cosmic shear dataset to evaluate its performance and to extract cosmological constraints across different physical scales.

\gsm{We find that the cosmological constraints remain broadly stable across the scale cuts considered in this \gsm{work. The largest difference between different physical-scale selections reaches $1.80\sigma$ under the ODC scenario, while the} resulting constraints remain consistent with the fiducial KiDS-Legacy findings} \cite{Wright2025KiDSLegacy,Stolzner25KiDSLegacy}\gsm{. We therefore find no evidence for a large modelling bias or scale-dependent systematic affecting} the current KiDS-Legacy cosmic shear analysis.
 
Nevertheless, the BNT scale-cut analysis reveals a mild but noteworthy dependence on the covariance prescription. When the observational data vector is used both in the likelihood and in the construction of the Gaussian part of the covariance matrix (ODC), the \gsm{largest difference of $1.80\sigma$} is found between the scale selection of $k < 0.33~\mathrm{Mpc}^{-1}$ and $k \in (0.33,3.3)~\mathrm{Mpc}^{-1}$\gsm{. More importantly, the ODC posteriors exhibit a coherent scale-dependent trend, with larger physical scales preferring lower $S_8$. This coherent behaviour across the scale selections makes the trend} useful as a diagnostic of the inference pipeline.
 
The interpretation of this trend changes when the theoretical-data-vector (TDC) covariance matrix is used. For TDC posteriors, $S_8$ constraints of different scale bins are more stable across scales, and the large-scale preference for lower $S_8$ is reduced to $0.75\sigma$. Mock tests using the theoretical prediction as the mock observational data vector show almost no difference between the ODC and TDC posterior results. This indicates that the data vector alone or the covariance matrix alone either cannot generate the large-scale deviation seen in the ODC analysis of the KiDS-Legacy pseudo-$C_\ell$ data vector. Instead, the apparent trend appears only when the observational pseudo-$C_\ell$ data vector and the observational-data-vector-based Gaussian covariance matrix are used together.

This result, therefore, should not be interpreted as evidence for a departure from the standard $\Lambda$CDM model. Although the ODC posterior is counterintuitive from the perspective of many theoretical attempts to resolve the $S_8$ tension through late-time or small-scale modifications, the TDC posteriors, as a comparison, show that the effect is mostly sensitive to the covariance prescription used in the likelihood. Our work shows that neither the data vector nor the covariance matrix is the sole origin of the effect; the trend appears when observational data vector is used simultaneously for covariance matrix calculation and likelihood computation. This suggests that our BNT-based $k$-cut method provides a sensitive test of the consistency as noted in our previous paper, \cite{Gu26BNT2}.

A possible explanation for the \gsm{behaviour} associated with the observational-data-vector covariance matrix is its treatment of the sample-variance, or cosmic-variance, contribution. If the actual observed spectrum is used in this part of the covariance calculation, the covariance matrix is no longer independent of the data vector that it is meant to weight. This \gsm{provides a possible mechanism for the data-covariance interplay seen in our results, particularly on large scales where cosmic variance is more important. Determining whether this mechanism is responsible for the observed trend requires a dedicated analysis beyond the tests presented here and is left to future work. In parallel, we are developing an ultrafast reconstruction method of the linear matter power spectrum. Its ultralow computational cost makes it well suited for additional validation of inference pipelines and for diagnosis of the covariance-prescription-dependent} behaviour observed in this KiDS-Legacy pseudo-$C_\ell$ analysis.

Although the statistical significance of the $S_8$ tension has been reduced by recent improvements in photometric-redshift calibration \cite{Wright2025KiDSLegacy}, scale-dependent tests remain important for investigating theoretical extensions to $\Lambda$CDM and for validating the robustness of weak-lensing inference pipelines. In this work, our results demonstrate that the BNT methodology is not only useful for isolating physical scales, but also for diagnosing theoretical-modelling systematics, data-vector characteristics, and covariance prescriptions. \gsm{More generally, the BNT-based $k$-cut framework provides a controlled environment for jointly testing cosmological information, observational systematics, and inference assumptions across physical scales.}
 
\begin{acknowledgments}
All theoretical calculations are based on \textsc{pyccl} \cite{Pyccl} and we present our whole sampling code suite in \cite{BNT_repo}. 
We calculate error bars and plot posterior contours with the \textsc{GetDist} \cite{getdist} package. We are grateful to Gary Bernstein, S\'ebastien Fabbro, Sacha Guerrini, Alan Heavens, Gary Hinshaw, Mike Hudson, Martin Kilbinger, Meng-Xiang Lin, Will Percival, Levon Pogosian, and the whole German Centre for Cosmological Lensing (GCCL) for their constructive feedback during the construction of the entire structure for the study and the analysis. S. G. acknowledges support from the John I. Watters Research Fellowship (Award No. 6701), and the Ruhr-Universit\"at Bochum Research Scholarship for Doctoral Researchers from International Partners. S. G. also thanks the Herzberg Astronomy and Astrophysics Research Centre of the Canadian National Research Council for hosting him during the finalisation phase of this work. Z. Y. is supported by JSPS KAKENHI Grant No. 23H00108 and JST FOREST Program Grant No. JPMJFR2137. L. v W. acknowledges support from the Natural Sciences and Engineering Research Council of Canada (NSERC) and the Canadian Institute for Advanced Research (CIFAR). H. H. is supported by a DFG Heisenberg Grant (No. Hi 1495/5-1), the DFG Collaborative Research Center SFB1491, an ERC Consolidator Grant (No. 770935), and the DLR Project No. 50QE2305. A. H. W. is supported by the Deutsches Zentrum f\"ur Luft- und Raumfahrt (DLR), under Project No. 50QE2305, made possible by the Bundesministerium f\"ur Wirtschaft und Klimaschutz, and acknowledges funding from the German Science Foundation DFG, via the Collaborative Research Center SFB1491 ``Cosmic Interacting Matters---From Source to Signal.'' M. B. is supported by the Polish National Science Center through Grant No. 2020/38/E/ST9/00395 and by the Polish Ministry of Science and Higher Education under the Grant Agreement No. 2026/WK/06. C. G. is funded by the MICINN Project No. PID2022-141079NB-C32. IFAE is partially funded by the CERCA programme of the Generalitat de Catalunya. S.-S. L. has received funding from the programme ``Netzwerke 2021,'' an initiative of the Ministry of Culture and Science of the State of Northrhine Westphalia. L. L. is supported by the Austrian Science Fund (FWF) [No. ESP 357-N] and funded in part by the Austrian Science Fund (FWF) 10.55776/F101300. L. M. acknowledges the financial contribution from the grant ASI No. 2024-10-HH.0 ``Attivit\`a scientifiche per la missione Euclid---fase E.'' R. R. is partially supported by an ERC Consolidator Grant (No. 770935). B. S. acknowledges support from the Max Planck Society and the Alexander von Humboldt Foundation in the framework of the Max Planck-Humboldt Research Award endowed by the Federal Ministry of Education and Research.
\end{acknowledgments}

\section{Data availability}
The code suite that used to compute the likelihoods is available at \cite{BNT_repo}, where different branches goes for different papers.

\bibliography{bnt_ia}% Produces the bibliography via BibTeX.
%\printbibliography

\end{document}
%
% ****** End of file apssamp.tex ******